# Wind turbine power and land cover effects on cumulative bat deaths

Aristides Moustakas[1,2,*], Panagiotis Georgiakakis[2], Elzbieta Kret[3], Eleftherios Kapsalis[4]

1. Infometrics Data Analytics Ltd, 128 City Road, London, EC1V 2NX, UK
2. Natural History Museum of Crete, University of Crete, Heraklion, Greece
3. WWF Greece, Charilaou Trikoupi 119-124, Athens, GR 114 73, Greece
4. Society for the Protection of Biodiversity of Thrace, Dadia, GR 684 00, Soufli, Greece

*Corresponding author,
Aris Moustakas
Email: arismoustakas@gmail.com

**Abstract**

Wind turbines (WT) cause bird and bat mortalities which depend on the WT and landscape features. The effects of WT features and environmental variables at different spatial scales associated to bat deaths in a mountainous and forested area in Thrace, NE Greece were investigated. Initially, we sought to quantify the most lethal WT characteristic between tower height, rotor diameter and power. The scale of interaction distance between bat deaths and the land cover characteristics surrounding the WTs was quantified. A statistical model was trained and validated against bat deaths and WT, land cover and topography features. Variance partitioning between bat deaths and the explanatory covariates was conducted. The trained model was used to predict bat deaths attributed to existing and future wind farm development in the region. Results indicated that the optimal interaction distance between WT and surrounding land cover was 5 km, the larger distance than the ones examined. WT power, natural land cover type and distance from water explained 40 %, 15 % and 11 % respectively of the total variance in bat deaths by WTs. The model predicted that operating but not surveyed WTs comprise of 377.8% and licensed but not operating yet will contribute to 210.2% additional deaths than the ones recorded. Results indicate that among all WT features and land cover characteristics, wind turbine power is the most significant factor associated to bat deaths. Results indicated that WTs located within 5 km buffer comprised of natural land cover types have substantial higher deaths. More WT power will result in more deaths. Wind turbines should not be licensed in areas where natural land cover at a radius of 5km exceeds 50%. These results are discussed in the climate-land use-biodiversity-energy nexus.

**Keywords**

Data analytics; wind turbine capacity; turbine power; bat fatalities; wind energy facilities; land use

**Introduction**

The world is changing subject to land use and climate changes (Almond et al., 2022; IPBES, 2019). In an effort to mitigate anthropogenic carbon emissions related to energy production, wind as renewable energy resources is considered a plausible alternative. However, installing wind power facilities is directly related to land loss and habitat fragmentation, through land transformation and roads (Kati et al., 2021). In addition, the operation of wind turbines is considered as additional threat to many protected wildlife species (Cryan et al., 2014). We are thus facing a paradox of impacting biodiversity to modulate anthropogenic climatic impacts, as there are strong interaction effects between biodiversity, land use and climate (Peters et al., 2019; Ritchie et al., 2020).

Wind energy facilities are increasing worldwide (Global Wind Energy Council, 2021). After major investment efforts in recent years, in 2021 Greece had already reached 44% of the 2030 national target for wind harnessing (Kati et al., 2021). The country is an attractive target for wind energy investments as it exhibits high wind potential combined with very high percentage of public land, of which the vast majority (>77%) is forests and grasslands (Spanos et al., 2015) and a supportive national climate policy for RES deployment (Kati et al., 2021). Greece is a global biodiversity hotspot (Myers et al., 2000) with a high number of plant and animal species as well as endemism. Natura 2000, the network of protected areas in European Union, simultaneously protects habitats, animal, and plant species (OECD, 2020). In Greece, the Natura 2000 network consists of 446 sites, which cover approximately 28% of the country's land and 20% of the marine surface area (https://www.wwf.gr/en/our_work/nature/marine/protected_areas/natura_2000/).
However, in Greece, wind energy facility development is permitted within Natura 2000 sites. Thus, the potential impacts of wind turbines (WT) merit detailed risk analysis as it might significantly contribute to habitat loss, fragmentation, and species decline.

Bats are long-lived mammals with low reproductive potential and thus high levels of adult individuals survivorship is important for maintaining a viable population (Wilkinson and South, 2002). All European bat species are strictly protected under the Habitats Directive (92/43/EEC, Annex IV) and the national legislation of many countries. Nevertheless, in the absence of specific, mandatory regulations, the effect of wind farm installation and operation on bat populations is rarely examined to a sufficient extent in environmental impact assessments, although increased mortality in wind turbines may increase extinction risk for frequently killed species (Frick et al., 2017).

Bat deaths depend upon characteristics of the environment around WTs such as cover of natural, agricultural, or artificial surface areas (Arnett and Baerwald, 2013; Arnett et al., 2008; Barré et al., 2023; Thompson et al., 2017). The interaction scale between bat abundance/deaths and the environmental characteristics varies between studies (Hartmann et al., 2021; Lehnert et al., 2014; Starbuck et al., 2022). In addition, the percentage of cover of each habitat type, the availability of water, trees, as well as topography is scale-specific (Starbuck et al., 2022). Furthermore, deaths depend on installed WT power (capacity) (MacGregor and Lemaître, 2020) and WT characteristics such as rotor diameter, speed, tower height, power, or produced energy that are correlated (Hartmann et al., 2021; Huso et al., 2021; MacGregor and Lemaître, 2020). Wind speed is a highly variable factor that significantly affects bat causalities in wind farms. Restricting nocturnal operation of WT to wind speeds above 5.0 to 6.0 m/s during the high risk periods has been proved to reduce dramatically the number of bat deaths with a minimum loss in produced energy (Arnett et al., 2011; Behr et al., 2017; Martin et al., 2017; Wellig et al., 2018).

Prediction is an elementary property of science (Medawar, 1984), yet ecology has traditionally focused on explaining patterns, rather than predicting beyond the range of the data, though this is changing (Evans et al., 2013). With modern online data repositories, censuring and monitoring methods such as remote sensing and acoustic or visual sensors, environmental data are more ubiquitous and richer than ever (Liu et al., 2021; Moustakas and Katsanevakis, 2018). In addition, modern computing capabilities and statistical methods facilitate the analysis of such data (Ma et al., 2017; Moustakas and Katsanevakis, 2018). More often than not, environmental data related to characteristics around the area of species of interest are easier to acquire than biological data, data of species *per se* (Daliakopoulos et al., 2017). Assuming that near features are more likely to be related than distant ones (Tobler, 1970), a locally trained model can be used to forecast biological patterns based on environmental and biological data in neighbouring locations (Daliakopoulos et al., 2017; Groom et al., 2018).

The use of statistical methods and modelling is of particular importance for the estimation of bats killed under WTs which might be often underestimated due to the fact that carcasses are of small size, not easily detected, quickly consumed by scavengers or destroyed from collisions (Smallwood, 2013). Death counts are improved by the use of detection dogs, cameras and digital technologies (Cryan et al., 2014; Smallwood et al., 2020), but these methods are costly and the scale and extent of their applications limited. A variety of methods have been tested to estimate bat mortalities, fatalities, or deaths based on existing data, including correction factors for scavenger removal, partial search of the WT surroundings and searcher efficiency (Arnett et al., 2005; Huso, 2011; Smallwood, 2013; Zimmerling and Francis, 2016), Stochastic Dynamic Methodology (Bastos et al., 2013), Bayes' theorem (Huso et al., 2015), habitat suitability modelling (Bastos et al., 2013) as well as complex statistical models and software tools (Maurer et al., 2020). Prediction of the cumulative effects of wind industry on a broader geographic scale has also challenged bat conservationists (Hayes, 2013; Roscioni et al., 2013; Voigt et al., 2012; Zimmerling and Francis, 2016).

In Greece, searches for bat deaths at wind farms are rarely undertaken and are based on variable search intensity and survey methodologies, in the absence of mandatory detailed guidelines. In one of the few studies, the results of intensive carcass searches in NE Greece were analysed with respect to species composition and temporal variation, but no inferences were made regarding WT features or environmental drivers of bat fatalities (Georgiakakis et al. 2012). In addition, surveys for killed bats in all neighbouring already operating wind energy facilities have not been conducted or had insufficient quality.

In this study we focused on modelling deaths at a set of WTs then predicting cumulative deaths across a larger region in a mountainous and forested area in Thrace, North East Greece. The main aims were: a) identification of WT and environmental features at different spatial scales associated to bat deaths, b) determine the most lethal WT feature, c) prediction of cumulative deaths in existing and future facilities, d) proposals for proper WT site selection related to landscape planning.

**Methods**

*Study area*

The study area is located in the Rhodope and Evros Regional Units, Thrace, Northern Greece. It is characterized as mountainous of rugged relief with slopes varying from gentle to steep (Vasilakis et al., 2008) and it is sparsely populated with scattered small villages and settlements. Altitude in the study area range from 659 m to 1020 m. Land cover is diverse including oak, junipers and endemic pine forests, as well as dry grasslands. Agricultural and artificial areas include mainly croplands and urban land covers. The climate is Mediterranean characterized by cold winters and mild/hot summers. Precipitation in winter ranges between 500 – 800 mm while in the summer is often < 20 mm. The study area is one of the richest in Europe in bat diversity, with at least 24 species present (Papadatou, 2010). Furthermore, it has exceptional ornithological importance as it hosts habitats that are of European-wide significance, mainly for large birds of prey and aquatic birds (Catsadorakis and Källander, 2010). A large part of the region has been selected as priority area for the development of wind energy, as it is also one of the areas with the highest wind capacity in mainland Greece (EuropeanCommission, 2022). The highest values of wind speed are observed in higher altitudes as well as in the coastal zone (RAE, 2022).

*Bat death data*

Bat death data collected between August 2009 to July 2010 in the frame of Worldwide Wildlife Fund (WWF) Greece research project were used (Georgiakakis et al., 2012). We use the term 'death' instead of 'mortality' or 'fatality' since the last two refer to percentages calculated as the number of deaths relative to the population size or dying per WT, both unknown to us. , that are both unknown to us. During that period 88 WT from 9 wind farms were surveyed for bat carcasses. Each WT was visited five or six days a week, except the period between late December 2009 and early March 2010 where visits were rarer due to weather – related restrictions in accessibility (Georgiakakis et al., 2012). In total each WT was visited 305 days within a year (26,840 sampling events) and the number of bat deaths (N = 167 individuals from nine species) was recorded. In cases of multiple deaths in the same WT and day, each death would be recorded as a separate data entry (line in the data) so that data are always in a binary form (1 = death, 0 = no death). The sampling approach did not allow the use of estimators to assess the probability of detection and correct for searcher efficiency, scavenging and other sources of underestimation. Therefore, only the raw count data were used in the analyses.

*WT data*

Spatial data (Fig. 1a) as well as technical characteristics in terms of rotor diameter (m), tower height (m) and power (kW) for each examined WT were retrieved from Regulatory Authority for Energy (RAE) geoportal (RAE, 2022). Power is used *sensu* nameplate capacity of each WT as given by the manufacturer, and it is not identical with the energy produced. Three layers were downloaded (operation, installation and production licenses respectively) as shape files from RAE geospatial map and clipped in Evros and Rhodope region borders. Obtained data were grouped in five datasets. Dataset 1 follows the 88 operating WTs where bat deaths were recorded in detail 2009-2010 (Georgiakakis et al., 2012); (Table 1). Datasets 2 and 3 consist of fully operational WTs poorly surveyed (other studies) and not surveyed at all for bat deaths, comprised of 38 and 139 WTs respectively (Table 1). Dataset 4 consists of 11 WTs that have an installation, but not an operation license yet (have been installed or will be soon, but their rotors are not yet spinning) (Table 1). Dataset 5 consists of 56 WTs under the production license, but without an installation or operation license yet (their power, location and technical characteristics are approved, but may not be installed for various reasons, after excluding those without Environmental Terms Approval); (Table 1).

*Environmental data & interaction scale*

We estimated the percentage of each land cover class around each WT at 250, 1000, and 5000 m radius in Quantum GIS to inspect the effect of habitat composition on bat deaths (Fig. 1b). For each radius, classification was performed using the most detailed resolution available of the CORINE land cover (CLC2018, level 3, spatial resolution 100 m); (EEA, 2010), to record the percentage of every land cover (%) class around each WT. Distance from the nearest trees (m) and distance from water (m) as the minimum distance between the WT and the land cover class containing trees and water were recorded. In addition, tree canopy cover (%) was calculated as the sum of all land cover types containing forest trees. Elevation (m), aspect (degrees), and slope (degrees) around each WT were also recorded. All variables explored in the analysis are listed in Table 2.

Climatic data are publicly available at a resolution of 1 km (Worldclim, 2013). However, they were not used due to the fact that WTs in the same farm would have identical climatic values, and WTs in different farms distancing less than 1.42 km in a straight line (the diagonal of 1 km$^2$ cell) would also get identical values.

*Selecting the land cover scale around WTs best explaining deaths*

Bat deaths occurring at the location of WT depend upon characteristics of the environment nearby such as natural, agricultural, or artificial habitats around WTs (Roeleke et al., 2016; Santos et al., 2013). The interaction scale between bat abundance/deaths and the environmental characteristics varies between studies (Lehnert et al., 2014; Santos et al., 2013; Starbuck et al., 2022). In addition, the percentage of cover of each habitat type, or the availability of water, trees as well as the mean topography is scale-specific (Starbuck et al., 2022).

To quantify the optimal interaction scale between bat deaths and land cover characteristics around WTs, a data-driven approach was used (Moustakas et al., 2019) by comparing models between deaths and land cover classes across radii or 250, 1000, and 5000 m. Three generalized linear models (GLM) were fit with bat death events (0, 1) per day per WT (dependent binary variable) and land cover type percentage (independent variables) at 250, 1000, and 5000 m, in the absence of any other variables. Land cover data included the Corine level 3 (most detailed) land cover classes (Table 2). Models were fit with a binomial family error structure to account for the binary nature of the dependent variable. Model selection was performed using the Akaike Information Criterion (AIC); (Akaike, 1974; Burnham and Anderson, 2002). Analysis was conducted using the 'lme4' package in R statistical software (R Development Core Team, 2022).

*Selecting the WT feature best explaining deaths*

To quantify the WT feature that is best fitted with bat deaths (Huso et al., 2021; Smallwood, 2013), the correlation between the available WT features, rotor diameter, tower height and power was calculated using a Pearson's correlation matrix. We sequentially performed data-driven selection of the WT feature better fitted with bat deaths: three GLMs were fit with bat death events per day per WT and power or tower height, or rotor diameter were used as independent explanatory variable in the absence of any other variables. Models were fit with a binomial family error structure and model selection was performed by using the AIC. Analysis was conducted using the 'lme4' package in R (R Development Core Team, 2022).

*WT and land cover effects on bat deaths*

A GLM was built with bat death events per day per WT (dependent variable) at the optimal interaction scale, and the WT feature better fitted with deaths, distance from trees, distance from water, elevation, aspect, slope, forest cover, artificial land cover, agricultural land cover and natural land cover percentage (independent variables). All natural, agricultural, and artificial Corine land cover types (listed in Table 2) were combined into a single natural, agricultural, and artificial land cover percentage variable respectively which was further used in the analysis. The model was fit with a binomial family error structure. All independent variables were log(x+1) transformed prior to the analysis to successfully account for heteroscedasticity, as well as for facilitating comparisons among variables of differing levels of magnitude. Model selection of the most parsimonious model structure containing the most informative variables only was performed by the AIC (Burnham and Anderson, 2002). Any deletion of non-significant variables that did not increase AIC > 2 was deemed justified (Burnham and Anderson, 2002). The output of the model is the probability that a fatality will be found predicted by the independent variables. Sequentially, model outputs were categorized into two groups based on their predicted probabilities (p) of death. Cases with p above or equal to 0.5 are considered as deaths, while < 0.5 as not (Kassambara, 2018). The sum of deaths across all outputs indicates total deaths across the 88 WT in one year. Model outputs were compared with data to quantify the deviance in model fitting. Analysis was conducted using the 'lme4', 'nlme' 'MuMIn', 'effects' 'jtools', 'ggplot2', 'sjmisc' and 'lattice' packages in R (R Development Core Team, 2022).

*Variance partitioning of variables explaining bat deaths*

Hierarchical variance partitioning was performed between the independent variables of the optimal GLM and bat deaths with a binomial regressor to account for the contribution of each explanatory variable to the total variance (Chevan and Sutherland, 1991). Variance partitioning is a computational statistical technique capable of handling potentially correlated independent variables, whilst ranking the predictor importance of each variable (Chevan and Sutherland, 1991; Mac Nally, 2002). Variance partitioning is calculated from the AIC weights of each independent variable and based upon the number of times that a variable was significant among all possible combinations of the explanatory variables (Mac Nally, 2002). Results add to 100%. We performed 99 randomizations to extract confidence intervals and significance of variables. Analysis was conducted using the 'hier.part' package in R (R Development Core Team, 2022).

*Predicting bat deaths*

The most parsimonious GLM between bat deaths and the explanatory independent variables was used for predicting bat deaths into datasets 2 – 5, where actual bat deaths are unknown, but the independent explanatory variables are available (Cameron and Trivedi, 2013; Chowell et al., 2020; Currie, 2016). The GLM input consisted of the values of each independent variable corresponding to the WT deaths to be predicted. The output of the model is death probability predicted by the independent variables. Sequentially, model outputs were categorized into two groups based on their predicted probabilities (p) of death. Cases with p above or equal to 0.5 are considered as deaths, while < 0.5 as not. The sum of deaths across all outputs indicates total deaths across all WTs in datasets 2 – 5, in a year. Analysis was conducted using the 'lme4' package in R (R Development Core Team, 2022).

**Results**

WT features were highly positively correlated as indicated by Pearson's correlation coefficient ($Corr_P$ = 0.846 between power – tower height; $Corr_P$ = 0.99 between power – rotor diameter; $Corr_P$ = 0.848 between tower height – rotor diameter, all p-values <<0.001). The best model fit between deaths and WT features was power as quantified by the lowest AIC score, followed by rotor diameter, while tower height ranked last from the three available WT features (Table 3). The lowest AIC model (deaths - power) significantly differed from the second best (deaths – rotor diameter); ANOVA, deviance= -4.53, p-value <0.01.

The best model fit between deaths and land cover radius was at 5000 m as indicated by the lowest AIC score of the model at this scale, followed by 250 m, while the model containing land cover classes percentages at 1000 m ranked third (Table 4). The lowest AIC model (radius 5000 m) significantly differed from the second best (radius 250 m); ANOVA, deviance =40.97, p-value <<0.001.

The most parsimonious (final) model included the effects of power, distance from trees and water, aspect, slope, agricultural land cover percentage and natural land cover percentage (Table 6). The inclusion of elevation, forest canopy cover percentage, and artificial land cover percentage were not justified by model selection and were thereby removed from the initial maximal model. The final model did not significantly differ from the initial maximal one, it was just simpler; ANOVA deviance = -0.947, p-value = 0.331 (Table 5). The mean of Artificial land cover percentage contained mainly zero values at the scale of 5000 m deployed in the final model, while elevation was highly correlated with slope, and forest canopy cover with natural land cover percentage (results not shown here).

In terms of model coefficients there is a negative fixed coefficient (-111.659) and the highest coefficient is fitted for the percentage of natural land within 5000 meters from each WT (mean coefficient = +17.345; Table 6, Fig. 2). The second and third highest coefficients were fit for distance from water (mean coefficient = +1.780) and power (mean coefficient = +1.260) respectively (Table 6, Fig. 2). The largest negative coefficient is fitted for slope (mean coefficient = -1.163; Table 6, Fig. 2). Sensitivity analysis of coefficients indicated that the effects of power, distance from water, slope and natural cover are robust as coefficients did not cross zero at a 95% confidence interval (Fig. 3).

Sensitivity analysis between the optimal model (after model selection) and the initial maximal one containing all variables indicated that 95% confidence intervals of coefficients of all three eliminated variables were crossing zero (results not shown here). In addition, 95% intervals of coefficients of variables included in the most parsimonious model and coefficients of the same variables before model selection did not significantly differ (results not shown here). We therefore did not proceed with model averaging across models and variables.

Variance partitioning indicated that power explained 40.22% of the total variance, the natural land cover percentage in 5000 m radius around each WT 15.93% and distance from water 11.52% (Figure 4). Slope explained 10.81%, agricultural cover 8.38%, aspect 8.22%, and distance from trees 4.88% of the total variance (Fig. 4). Randomizations indicated that all variables were significant (Table 7).

The model predicted 169 deaths in dataset 1 when the actual recorded deaths were 167, and thus overestimated deaths by 1.2% (Fig. 5). Model predictions in dataset 2 indicated 152 deaths, 479 deaths in dataset 3, 79 deaths in data set 4, and 272 deaths in dataset 5 per year (Fig. 5). The total predicted number of bat deaths in the wind turbines of the study area is 1151 bats per year. In terms of percentages in comparison with recorded deaths (dataset 1), there is a higher probability of deaths of 377.8% in operating, but not sufficiently surveyed WTs (dataset 2 & 3). WTs not operating yet, but bound to (data set 4) or likely to operate soon (data set 5) together will contribute to 210.2% excess deaths than the ones recorded. Under the scenario that the WTs under production and installation licence proceed to operation, the total excess deaths will be 588% higher than the ones recorded.

**Discussion**

*Deaths and wind turbine power*

Results indicate deaths are significantly associated to wind turbine power. Power alone explained a third of the variance of bat deaths and it is more lethal than tower height or rotor diameter. Energy, the product of power and time, produced by WT is determined by the size of the turbine (tower height, rotor diameter), and rotor speed (Dixon and Hall, 2014). To that end decreasing rotor speed would need to be compensated by increasing WT size and *vice versa* for producing the same amount of energy. Thus, altering WT features will not minimize impacts on bats, as long as the energy produced remains the same or even increases. In addition, having several small or fewer large WT (Dabiri et al., 2015; Krijgsveld et al., 2009) is unlikely to minimize deaths as long as total produced energy remains constant. Avian and bat mortality was reported to be constant per energy unit produced across wind turbine power capacities (Huso et al., 2021), confirming that not simply the size of turbines but power is the factor better explaining deaths. In the same study, the actual energy produced by each WT, but not the nominal power which is the maximum power capacity was found as a more unbiased estimator of wildlife deaths (Huso et al., 2021). Since the actual energy produced by turbines in each wind farm is not disclosed in Greece, nominal WT power is still a better predictor of deaths than other WT features.

Another study also found power to be a significant explanatory variable with more installed capacity correlating with higher deaths (MacGregor and Lemaître, 2020). However the same study suggest that power 'was a poor predictor of estimated mortality' (MacGregor and Lemaître, 2020). One explanation why we found power (capacity) to be a strong predictor could be that all of the sampled WT were operating very similarly. This is plausible, because at the study area WTs are located in mountain or hill summits and relatively closely located at a straight line. Thus, there are few barriers that can obstruct or de-synchronize wind patterns during the year. In this case energy production, turbine operation and power will be highly correlated. The spatial synchrony of climate is also known to synchronize animal populations, the Moran effect (Moran, 1953), and synchronized populations suffer higher casualties (Hansen et al., 2013; Moustakas et al., 2018).

*Land cover and the interaction zone*

Natural land cover was the second best predictor of bat deaths in terms of variance explained. It also had the steepest coefficient slope across all variables, deaths steeply increased for every unit (1%) of natural land cover increase. Thus, the 'perfect storm' for bat deaths needs a combination of high total power and WT installed in natural land cover areas. To that end our results confirm that 'the larger a facility is, the more important specific spatial and environmental context becomes in determining bat mortality' (MacGregor and Lemaître, 2020).

Deaths caused at operating WTs depend on land cover characteristics of coarser scales (Lehnert et al., 2014). Our results, in addition to previous studies (Dietz and Kiefer, 2016; Rodrigues et al., 2015b) suggest that in natural areas, wind turbines should not be licensed in areas where natural land cover at a radius of 5km exceeds 50%. . Natural land cover in terms of coniferous and broadleaved forests are associated with foraging areas and high mobility (Ferreira et al., 2015) which is reasonable to result in higher deaths. High bat collision rates with WTs are also reported in forested mountain ridge tops in the US (Kunz et al., 2007). . Installing of wind farms in disturbed areas instead of natural ones, is necessary in order to minimize cumulative impacts to wildlife (Kiesecker et al., 2011).

Outside these high-risk areas, environmental impacts assessment studies should extend the search range for important foraging, drinking and roosting sites in a sufficiently large area, typically no smaller than 5 km. The spatial cumulative effect of at least 5 km should be assessed as a combined effect of wind energy developments comprised of multiple WTs from different wind farms, taken together, rather just for a single WT. The interaction zone of at least or close to 5 km between the land cover and WT is deduced also from other studies: most processes regarding bat occupancy prediction were found to be most significant at 5760 m radius (Starbuck et al., 2022), the scale closest to 5 km from the ones examined in that study. Higher bat mortalities in WT located closer than 5 km from forests were also reported in Portugal (Santos et al., 2013), while Barré et al. (2023) found that bat activity around nacelles was positively affected by landscape characteristics at a radius of up to 10 km.

Other explanatory covariates such as agricultural land cover, aspect, slope, distance from trees and water yielded similar results with other studies (Ferreira et al., 2015; Roeleke et al., 2016; Roemer et al., 2019; Santos et al., 2013; Starbuck et al., 2022). The lack of significance of elevation can be explained by the fact that the data set is comprised of a mountainous landscape and thus there are few lower elevation values to allow for differentiating effects. It appears that in this case, slope is more important. Artificial land cover at a radius of 5 km around each WT was comprised in most cases of values very close to zero, and thus the lack of significance in the optimal model is based on the fact that WTs are located in a landscape that is mainly natural and agricultural at that scale.

*Monitoring and public data repositories*

In the absence of proper regulations for impact of WTs on bats in the post-construction monitoring procedure, deaths that pass virtually unnoticed are several folds higher than the actual recorded ones. As bats have low reproduction rates, this is alarming as their population viability (Frick et al., 2017) or extinction debt numbers (Chattopadhyay et al., 2019) might subtly exceed reproductive population thresholds. To that end automated monitoring (McClure et al., 2018) together with field surveys should be systematically regulated together with the operational license to avoid severe population declines and regional or national scale extinctions (Rodrigues et al., 2015b). There is a need of high accuracy when monitoring the wind farm impacts, especially at large sites with steep and heavily vegetated areas. The use of detection dogs to effectively monitor bird and bat deaths at wind farms is becoming increasingly popular (Bennett, 2015). All studies to date agree that dogs outperform human searchers at finding bird and bat carcasses around wind turbines (Domínguez del Valle et al., 2020). According to decisions for the approval of environmental terms designed by the Greek authorities, carcass search area should be extended to 400 m around a turbine and a buffer zone of 300 m from roads connecting wind turbines within important bird areas and special protection areas sites, what makes it extremely difficult to be achieved by human researchers, thus the use of dogs should be obligatory implemented. In addition, online publicly accessible dataset repositories on species deaths by wind turbines should be established (Fernández-Bellon, 2020), regulated by European Commission as well as at national level. There is a need for publicly available data compiled in a comparable format, facilitates' cumulative risk assessment across regions, informed democratic decisions and social awareness (Murray-Rust, 2008). Such data are already available among others for road kills (Balčiauskas et al., 2020; Englefield et al., 2020) or invasive alien species (Deriu et al., 2017) and data-driven research is used to inform policy (Vanderhoeven et al., 2017).

*Uncertainties and future research*

Deaths were best explained in an interaction zone of 5 km which is the larger of the scales examined here. Further studies expanding that scale are needed to investigate the optimal distance – see e.g. (Moustakas et al., 2019) for an application in trees. It is worth noting that the relationship between deaths, WTs and land cover is not linear as the finer scale, 250 m exhibited a better model fit than the intermediate scale of 1 km. The analysis did not differentiate among bat species as models could not be fitted for each species separately (Santos et al., 2013). In addition, the interaction zone of 5 km between land cover and WTs needs to be cross-validated with bird species. In terms of data analytics, model validation was performed by comparing outputs of the trained GLM against the data that were used for training. Data scarcity did not permit in splitting the data into training and testing, with testing data not used for training (de Hond et al., 2022). The inclusion of climatic data especially at temporal resolutions within the year (months, weeks, etc) could provide valuable insights regarding WT synchronization as well as bat synchronization (Nicolau et al., 2022).

The number of bat deaths recorded in the field used for model fitting is an underestimate of the actual one. Although there are statistical models for estimating WT bat deaths number e.g. GenEst (Simonis et al., 2018), the sampling design did not permit using such methods to estimate deaths from raw count data. Obviously, the actual number of bat deaths in the study area is higher, even in WTs where no carcasses were found (Arnett, 2005; Huso et al., 2015), but a precise estimation was not possible.

Furthermore, although not examined in our study, landscape configuration in terms of mountain passes, linear elements suitable for commuting between habitat fragments (e.g. treelines, hedgerows, ravines), and water bodies too small to be classified into the Corine land cover, should be taken in account during the planning and licensing process (Arnett et al., 2008; Kelm et al., 2014; Piorkowski and O'Connell, 2010; Rodrigues et al., 2015a). Other configuration features such as habitat complexity has to been taken into account as bat deaths at WTs are expected to be higher in structurally rich landscapes (Hurst et al., 2016).

*Landscape planning implications*

The target of the new EU biodiversity strategy is to increase the protected area to at least 30 % of EU land by 2030 (EuropeanCommission, 2021; Hurst et al., 2016). Primary and old-growth forests and other carbon-rich ecosystems, such as grasslands will be the focus of conservation efforts, as well as soil conservation (EuropeanCommission, 2021). Rewilding EU is also core part of the plan (EuropeanCommission, 2021). On top of this plan, a second pioneering proposal aims to repair the 80% of European habitats that are in poor condition, and to bring back nature to all ecosystems (EuropeanCommission, 2022).

Installing WTs in protected areas or old forests and causing species decline is clearly contradicting both strategic plans. WTs are directly related with land loss and degradation by artificial base surfaces and road-induced fragmentation. Ranking of biodiversity threats in Europe indicates that historically threats deriving from land use changes are several folds higher than the ones from climatic changes (Almond et al., 2020), with comparable results in other studies (IPBES, 2019). In the most recent assessments, land-use change is still the biggest current threat to nature, destroying or fragmenting the natural habitats of many plant and animal species on land, in freshwater and in the sea (Almond et al., 2022). That is not meant to downplay climate-derived threats, but rather to disentangle the impact of land use changes and climatic changes on ecosystems.

A paradox emerges of impacting biodiversity to mitigate the climate: biodiversity is negatively affected by climate change, but biodiversity, through the ecosystem services it supports, makes an important contribution to both climate mitigation and adaptation (EuropeanCommission, 2009). Consequently, conserving and sustainably managing biodiversity is critical to addressing climate change (EuropeanCommission, 2009; Lloret et al., 2022). It thus needs to be prioritized what is the relative gain and loss of wind farm installation and operation in the climate-land-biodiversity-energy nexus.

**Acknowledgements**

This work is a part of the Project "Safe Flyways- reducing energy infrastructure related bird mortality in the Mediterranean" founded by MAVA Foundation and BirdLife International. We thank Lavrentis Sidiropoulos for his support with GIS analysis, and several WWF volunteers for field work. Comments from two anonymous reviewers considerably improved an earlier manuscript draft.

**References:**

Akaike H. New look at statistical-model identification. IEEE Transactions on Automatic Control 1974; AC19: 716-723.

Almond REA, Grooten M, Juffe Bignoli D, Petersen T. Living Planet Report 2022 – Building a naturepositive society. WWF, Gland, Switzerland, 2022.

Almond REA, Grooten M, Peterson T. Living Planet Report 2020-Bending the curve of biodiversity loss. WWF, Gland, Switzerland: World Wildlife Fund, 2020.

Arnett EB, Baerwald EF. Impacts of wind energy development on bats: implications for conservation. Bat evolution, ecology, and conservation 2013: 435-456.

Arnett EB, Brown WK, Erickson WP, Fiedler JK, Hamilton BL, Henry TH, et al. Patterns of bat fatalities at wind energy facilities in North America. The Journal of Wildlife Management 2008; 72: 61-78.

Arnett EB, Erickson WP, Kerns J, Horn J. Relationships between bats and wind turbines in Pennsylvania and West Virginia: an assessment of bat fatality search protocols, patterns of fatality, and behavioral interactions with wind turbines. A final report submitted to the bats and wind energy cooperative. Bat Conservation International, Austin, Texas, USA 2005.

Arnett EB, Huso MM, Schirmacher MR, Hayes JP. Altering turbine speed reduces bat mortality at wind‐energy facilities. Frontiers in Ecology and the Environment 2011; 9: 209-214.

Balčiauskas L, Stratford J, Balčiauskienė L, Kučas A. Importance of professional roadkill data in assessing diversity of mammal roadkills. Transportation Research Part D: Transport and Environment 2020; 87: 102493.

Barré K, Froidevaux JS, Sotillo A, Roemer C, Kerbiriou C. Drivers of bat activity at wind turbines advocate for mitigating bat exposure using multicriteria algorithm-based curtailment. Science of The Total Environment 2023: 161404.

Bastos R, Santos M, Cabral JA. A new stochastic dynamic tool to improve the accuracy of mortality estimates for bats killed at wind farms. Ecological indicators 2013; 34: 428-440.

Behr O, Brinkmann R, Hochradel K, Mages J, Korner-Nievergelt F, Niermann I, et al. Mitigating bat mortality with turbine-specific curtailment algorithms: A model based approach. Wind Energy and Wildlife Interactions: Presentations from the CWW2015 Conference. Springer, 2017, pp. 135-160.

Bennett E. Observations from the use of dogs to undertake carcass searches at wind facilities in Australia. Wind and wildlife. Springer, 2015, pp. 113-123.

Burnham KP, Anderson DR. Model Selection and Multimodel Inference. New York: Springer Verlag, 2002.

Cameron AC, Trivedi PK. Regression analysis of count data. Vol 53: Cambridge university press, 2013.

Catsadorakis G, Källander H. The Dadia–Lefkimi–Soufli Forest National Park, Greece: Biodiversity, Management and Conservation. WWF Greece, Athens, pp. 215–226. 2010.

Chattopadhyay B, Garg KM, Mendenhall IH, Rheindt FE. Historic DNA reveals Anthropocene threat to a tropical urban fruit bat. Current Biology 2019; 29: R1299-R1300.

Chevan A, Sutherland M. Hierarchical partitioning. The American Statistician 1991; 45: 90-96.

Chowell G, Luo R, Sun K, Roosa K, Tariq A, Viboud C. Real-time forecasting of epidemic trajectories using computational dynamic ensembles. Epidemics 2020; 30: 100379.

Cryan PM, Gorresen PM, Hein CD, Schirmacher MR, Diehl RH, Huso MM, et al. Behavior of bats at wind turbines. Proceedings of the National Academy of Sciences 2014; 111: 15126-15131.
Currie ID. On fitting generalized linear and non-linear models of mortality. Scandinavian Actuarial Journal 2016; 2016: 356-383.
Dabiri JO, Greer JR, Koseff JR, Moin P, Peng J. A new approach to wind energy: Opportunities and challenges. AIP Conference Proceedings 2015; 1652: 51-57.
Daliakopoulos IN, Katsanevakis S, Moustakas A. Spatial Downscaling of Alien Species Presences Using Machine Learning. Frontiers in Earth Science 2017; 5: 60.
de Hond AAH, Leeuwenberg AM, Hooft L, Kant IMJ, Nijman SWJ, van Os HJA, et al. Guidelines and quality criteria for artificial intelligence-based prediction models in healthcare: a scoping review. npj Digital Medicine 2022; 5: 2.
Deriu I, D'Amico F, Tsiamis K, Gervasini E, Cardoso AC. Handling Big Data of Alien Species in Europe: The European Alien Species Information Network Geodatabase. Frontiers in ICT 2017; 4: 20.
Dietz C, Kiefer A. Bats of Britain and Europe: Bloomsbury Publishing, 2016.
Dixon S, Hall C. Wind Turbines. Fluid Mechanics and Thermodynamics of Turbomachinery 2014: 419-485.
Domínguez del Valle J, Cervantes Peralta F, Jaquero Arjona MI. Factors affecting carcass detection at wind farms using dogs and human searchers. Journal of Applied Ecology 2020; 57: 1926-1935.
EEA. CORINE land cover 2010. European Environment Agency 2010.
Englefield B, Starling M, Wilson B, Roder C, McGreevy P. The Australian Roadkill Reporting Project—Applying Integrated Professional Research and Citizen Science to Monitor and Mitigate Roadkill in Australia. Animals 2020; 10: 1112.
EuropeanCommission. WHITE PAPER: Adapting to climate change: Towards a European framework for action: Commission of the European Communities, 2009.
EuropeanCommission. EU biodiversity strategy for 2030 : bringing nature back into our lives: Publications Office of the European Union, 2021.
EuropeanCommission. Proposal for a REGULATION OF THE EUROPEAN PARLIAMENT AND OF THE COUNCIL on nature restoration. COM (2022) 304, 2022.
Evans MR, Bithell M, Cornell SJ, Dall SRX, Díaz S, Emmott S, et al. Predictive systems ecology. Proceedings of the Royal Society B: Biological Sciences 2013; 280: 20131452.
Fernández-Bellon D. Limited accessibility and bias in wildlife-wind energy knowledge: A bilingual systematic review of a globally distributed bird group. Science of The Total Environment 2020; 737: 140238.
Ferreira D, Freixo C, Cabral JA, Santos R, Santos M. Do habitat characteristics determine mortality risk for bats at wind farms? Modelling susceptible species activity patterns and anticipating possible mortality events. Ecological Informatics 2015; 28: 7-18.
Frick WF, Baerwald EF, Pollock JF, Barclay RM, Szymanski J, Weller T, et al. Fatalities at wind turbines may threaten population viability of a migratory bat. Biological Conservation 2017; 209: 172-177.
Georgiakakis P, Kret E, Cárcamo B, Doutau B, Kafkaletou-Diez A, Vasilakis D, et al. Bat fatalities at wind farms in north-eastern Greece. Acta Chiropterologica 2012; 14: 459-468.
Global Wind Energy Council G. Global wind report 2021. Global Wind Energy Council Brussels, Belgium, 2021.
Groom QJ, Marsh CJ, Gavish Y, Kunin WE. How to predict fine resolution occupancy from coarse occupancy data. Methods in Ecology and Evolution 2018; 9: 2273-2284.

Hansen BB, Grøtan V, Aanes R, Sæther B-E, Stien A, Fuglei E, et al. Climate Events Synchronize the Dynamics of a Resident Vertebrate Community in the High Arctic. Science 2013; 339: 313-315.

Hartmann SA, Hochradel K, Greule S, Günther F, Luedtke B, Schauer-Weisshahn H, et al. Collision risk of bats with small wind turbines: Worst-case scenarios near roosts, commuting and hunting structures. PLOS ONE 2021; 16: e0253782.

Hayes MA. Bats killed in large numbers at United States wind energy facilities. BioScience 2013; 63: 975-979.

Hurst J, Biedermann M, Dietz C, Dietz M, Karst I, Krannich E, et al. NaBiV Heft 153: Fledermäuse und Windkraft im Wald. Bundesamt für Naturschutz, 2016.

Huso M, Conkling T, Dalthorp D, Davis M, Smith H, Fesnock A, et al. Relative energy production determines effect of repowering on wildlife mortality at wind energy facilities. Journal of Applied Ecology 2021; 58: 1284-1290.

Huso MM. An estimator of wildlife fatality from observed carcasses. Environmetrics 2011; 22: 318-329.

Huso MM, Dalthorp D, Dail D, Madsen L. Estimating wind‐turbine‐caused bird and bat fatality when zero carcasses are observed. Ecological Applications 2015; 25: 1213-1225.

IPBES. Global assessment report on biodiversity and ecosystem services of the Intergovernmental Science-Policy Platform on Biodiversity and Ecosystem Services. In: Brondizio ES, Settele J, Díaz S, Ngo HT, editors, Bonn, Germany, 2019.

Kassambara A. Machine learning essentials: Practical guide in R: Sthda, 2018.

Kati V, Kassara C, Vrontisi Z, Moustakas A. The biodiversity-wind energy-land use nexus in a global biodiversity hotspot. Science of The Total Environment 2021; 768: 144471.

Kelm DH, Lenski J, Kelm V, Toelch U, Dziock F. Seasonal bat activity in relation to distance to hedgerows in an agricultural landscape in central Europe and implications for wind energy development. Acta Chiropterologica 2014; 16: 65-73.

Kiesecker JM, Evans JS, Fargione J, Doherty K, Foresman KR, Kunz TH, et al. Win-win for wind and wildlife: a vision to facilitate sustainable development. PLoS One 2011; 6: e17566.

Krijgsveld KL, Akershoek K, Schenk F, Dijk F, Dirksen S. Collision risk of birds with modern large wind turbines. Ardea 2009; 97: 357-366.

Kunz TH, Arnett EB, Cooper BM, Erickson WP, Larkin RP, Mabee T, et al. Assessing impacts of wind‐energy development on nocturnally active birds and bats: a guidance document. The Journal of Wildlife Management 2007; 71: 2449-2486.

Lehnert LS, Kramer-Schadt S, Schönborn S, Lindecke O, Niermann I, Voigt CC. Wind farm facilities in Germany kill noctule bats from near and far. PloS one 2014; 9: e103106.

Liu J, Xiang J, Jin Y, Liu R, Yan J, Wang L. Boost Precision Agriculture with Unmanned Aerial Vehicle Remote Sensing and Edge Intelligence: A Survey. Remote Sensing 2021; 13: 4387.

Lloret J, Turiel A, Solé J, Berdalet E, Sabatés A, Olivares A, et al. Unravelling the ecological impacts of large-scale offshore wind farms in the Mediterranean Sea. Science of The Total Environment 2022; 824: 153803.

Ma Z, Xie J, Li H, Sun Q, Si Z, Zhang J, et al. The role of data analysis in the development of intelligent energy networks. IEEE Network 2017; 31: 88-95.

Mac Nally R. Multiple regression and inference in ecology and conservation biology: further comments on identifying important predictor variables. Biodiversity & Conservation 2002; 11: 1397-1401.

MacGregor KA, Lemaître J. The management utility of large-scale environmental drivers of bat mortality at wind energy facilities: The effects of facility size, elevation and geographic location. Global Ecology and Conservation 2020; 21: e00871.

Martin CM, Arnett EB, Stevens RD, Wallace MC. Reducing bat fatalities at wind facilities while improving the economic efficiency of operational mitigation. Journal of Mammalogy 2017; 98: 378-385.

Maurer JD, Huso M, Dalthorp D, Madsen L, Fuentes C. Comparing methods to estimate the proportion of turbine-induced bird and bat mortality in the search area under a road and pad search protocol. Environmental and Ecological Statistics 2020; 27: 769-801.

McClure CJW, Martinson L, Allison TD. Automated monitoring for birds in flight: Proof of concept with eagles at a wind power facility. Biological Conservation 2018; 224: 26-33.

Medawar P. The Limits of Science. Oxford, UK: Oxford University Press, 1984.

Moran P. The statistical analysis of the Canadian Lynx cycle. Australian Journal of Zoology 1953; 1: 291-298.

Moustakas A, Daliakopoulos IN, Benton TG. Data-driven competitive facilitative tree interactions and their implications on nature-based solutions. Science of The Total Environment 2019; 651: 2269-2280.

Moustakas A, Evans MR, Daliakopoulos IN, Markonis Y. Abrupt events and population synchrony in the dynamics of Bovine Tuberculosis. Nature Communications 2018; 9: 2821.

Moustakas A, Katsanevakis S. Editorial: Data Mining and Methods for Early Detection, Horizon Scanning, Modelling, and Risk Assessment of Invasive Species. Frontiers in Applied Mathematics and Statistics 2018; 4.

Murray-Rust P. Open data in science. Nature Precedings 2008: 1-1.

Myers N, Mittermeier RA, Mittermeier CG, da Fonseca GAB, Kent J. Biodiversity hotspots for conservation priorities. Nature 2000; 403: 853-858.

Nicolau P, Ims R, Sørbye S, Yoccoz N. Seasonality, density dependence, and spatial population synchrony. Proceedings of the National Academy of Sciences of the United States of America 2022; 119: e2210144119.

OECD. OECD Environmental Performance Reviews: Greece 2020, 2020.

Papadatou E. Bats (Mammalia: Chiroptera): species diversity, distribution and abundance. – In: Catsadorakis, G. and Källander, H. (eds). The Dadia–Lefkimi–Soufli Forest National Park, Greece: Biodiversity, Management and Conservation. WWF Greece, Athens, pp. 215–226., 2010.

Peters MK, Hemp A, Appelhans T, Becker JN, Behler C, Classen A, et al. Climate–land-use interactions shape tropical mountain biodiversity and ecosystem functions. Nature 2019; 568: 88-92.

Piorkowski MD, O'Connell TJ. Spatial pattern of summer bat mortality from collisions with wind turbines in mixed-grass prairie. The American midland naturalist 2010; 164: 260-269.

R Development Core Team. R: A language and environment for statistical computing. R Foundation for Statistical Computing, Vienna, Austria. ISBN 3-900051-07-0. 2022.

RAE. Geospatial Map for energy units and requests, 2022.

Ritchie PD, Smith GS, Davis KJ, Fezzi C, Halleck-Vega S, Harper AB, et al. Shifts in national land use and food production in Great Britain after a climate tipping point. Nature Food 2020; 1: 76-83.

Rodrigues L, Bach L, Dubourg-Savage M-J, Karapandža B, Rnjak D, Kervyn T, et al. Guidelines for consideration of bats in wind farm projects. Eurobats Publication Series ISBN 978-92-95058-31-6, 2015a, pp. 1-51.

Rodrigues L, Bach L, Dubourg-Savage M-J, Karapandža B, Rnjak D, Kervyn T, et al. Guidelines for consideration of bats in wind farm projects Revision 2014, 2015b.

Roeleke M, Blohm T, Kramer-Schadt S, Yovel Y, Voigt CC. Habitat use of bats in relation to wind turbines revealed by GPS tracking. Scientific reports 2016; 6: 1-9.

Roemer C, Bas Y, Disca T, Coulon A. Influence of landscape and time of year on bat-wind turbines collision risks. Landscape Ecology 2019; 34: 2869-2881.

Roscioni F, Russo D, Di Febbraro M, Frate L, Carranza M, Loy A. Regional-scale modelling of the cumulative impact of wind farms on bats. Biodiversity and Conservation 2013; 22: 1821-1835.

Santos H, Rodrigues L, Jones G, Rebelo H. Using species distribution modelling to predict bat fatality risk at wind farms. Biological Conservation 2013; 157: 178-186.

Simonis J, Dalthorp D, Huso MM, Mintz J, Madsen L, Rabie PA, et al. GenEst user guide—Software for a generalized estimator of mortality. Techniques and Methods, Reston, VA, 2018.

Smallwood KS. Comparing bird and bat fatality-rate estimates among North American wind-energy projects. Wildlife Society Bulletin 2013; 37: 19-33.

Spanos I, Meliadis I, Platis P, Mantzanas K, Samara T, Meliadis M, et al. Forest Land Ownership Change in Greece. COST Action FP1201 FACESMAP Country Report, Vienna, 2015, pp. 31.

Starbuck CA, Dickson BG, Chambers CL. Informing wind energy development: Land cover and topography predict occupancy for Arizona bats. Plos one 2022; 17: e0268573.

Thompson M, Beston JA, Etterson M, Diffendorfer JE, Loss SR. Factors associated with bat mortality at wind energy facilities in the United States. Biological Conservation 2017; 215: 241-245.

Tobler WA. A Computer Movie Simulating Urban Growth in the Detroit Region. Economic Geography 1970; 46.

Vanderhoeven S, Adriaens T, Desmet P, Strubbe D, Barbier Y, Brosens D, et al. Tracking Invasive Alien Species (TrIAS): Building a data-driven framework to inform policy. Research Ideas and Outcomes 2017; 3: e13414.

Vasilakis DP, Poirazidis KS, Elorriaga JN. Range use of a Eurasian black vulture (Aegypius monachus) population in the Dadia‐Lefkimi‐Soufli National Park and the adjacent areas, Thrace, NE Greece. Journal of Natural History 2008; 42: 355-373.

Voigt CC, Popa-Lisseanu AG, Niermann I, Kramer-Schadt S. The catchment area of wind farms for European bats: a plea for international regulations. Biological conservation 2012; 153: 80-86.

Wellig SD, Nusslé S, Miltner D, Kohle O, Glaizot O, Braunisch V, et al. Mitigating the negative impacts of tall wind turbines on bats: Vertical activity profiles and relationships to wind speed. PloS one 2018; 13: e0192493.

Wilkinson GS, South JM. Life history, ecology and longevity in bats. Aging cell 2002; 1: 124-131.

Worldclim. Global Climate Data. Free climate data for ecological modeling and GIS, http://www.worldclim.org, 2013.

Zimmerling JR, Francis CM. Bat mortality due to wind turbines in Canada. The Journal of Wildlife Management 2016; 80: 1360-1369.

**Table 1**. Data sets used in the analysis. Data set 1: wind turbines surveyed for a full year (2009-2010) and used for model training and variance partitioning; data sets 2 & 3: wind turbines in operation, but not adequately surveyed; data sets 4 & 5: licensed, but not in operation yet. Datasets 2 – 5 were used for predicting bat deaths using the trained model in data set 1. The number of wind turbines (Nr of WT), total power (Sum of all WTs power; kW), and power per wind turbine (kW/WT) are listed in each data set.

| Dataset | Windfarm category | Nr of WT | Total power (kW) | Power range (kW) |
|---|---|---|---|---|
| 1 | Operational, thoroughly surveyed | 88 | 106300 | 800 - 2000 |
| 2 | Operational, poorly surveyed | 38 | 85600 | 2000 - 2500 |
| 3 | Operational, not surveyed | 139 | 270300 | 600 - 3600 |
| 4 | Licensed, not operating | 11 | 44400 | 3600 - 4200 |
| 5 | Licence pending | 56 | 153700 | 2000 - 4200 |

**Table 2.** Variables used in the analysis. Scale refers to the radius around each WT explored. WT features are the allometric and energy related characteristics of each WT. Topography refer to the physical landscape characteristics at the location of each WT and its distance to the nearest water sources and trees. Land cover follows the CORINE level 3 classification system. Land cover types are the ones that appeared in the classification around WTs. The sum of 'Broad leaved forest', 'Coniferous forest', and 'Mixed forest land cover' type percentages appears as the variable 'Forest land cover', that was also explored in the analysis

| Category | Variables | Units | Mean |
|---|---|---|---|
| **Scale** | | | |
| | 250 Radius | m | |
| | 1000 Radius | m | |
| | 5000 Radius | m | |
| **Wind Turbine** | | | |
| | Power | kW | 1208 |
| | Tower Height | m | 54.132 |
| | Rotor Diameter | m | 60.506 |
| **Topography** | | | |
| | Elevation | m | 867.95 |
| | Aspect | degrees | 172.61 |
| | Slope | degrees | 14.475 |
| | Distance from water | m | 18372 |
| | Distance from trees | m | 65.874 |
| **Land cover** | | | |
| *Artificial land cover* | | | |
| | Discontinues urban fabric | % | <0.001 |
| | Industrial or commercial units | % | <0.001 |
| | Road and rail network and associated land | % | <0.001 |
| | Mineral extraction sites | % | <0.001 |
| *Agricultural land cover* | | % | <0.001 |
| | Non-irrigated arable land | % | <0.001 |
| | Olive groves | % | <0.001 |
| | Pastures | % | <0.001 |
| | Complex cultivation patterns | % | 5.0815 |
| | Land principally occupied by agriculture, with significant areas of natural vegetation | % | 42.978 |
| *Natural land cover* | | | |
| | Broad leaved forest | % | 12.727 |
| | Coniferous forest | % | 0.449 |
| | Mixed forest | % | 10.388 |
| | Natural grassland | % | 25.591 |
| | Sclerophyllous vegetation | % | 2.3011 |
| | Transitional woodland scrub | % | 0.489 |
| | Sparsely vegetated areas | % | <0.001 |

| Water bodies | % | <0.001 |

**Table 3.** Statistics of the three generalized linear models used to select the wind turbine feature best explaining bat deaths.

| WT feature | Resid.df | Resid.Dev | Deviance | AIC |
|---|---|---|---|---|
| Power | 26749 | 2172.1 | | 2176.09 |
| Rotor diameter | 26749 | 2176.6 | -4.53 | 2180.61 |
| Tower height | 26749 | 2201.8 | -25.17 | 2205.78 |

**Table 4.** Statistics of the three generalized linear models used to select the land cover distance around wind turbines best explaining bat deaths.

| Radius  | Resid.df | Resid.Dev | Deviance | AIC     |
|---------|----------|-----------|----------|---------|
| 250 m   | 26744    | 2163.9    |          | 2177.95 |
| 1000 m  | 26744    | 2195.6    | -31.60   | 2209.55 |
| 5000 m  | 26742    | 2154.6    | 40.97    | 2172.58 |

**Table 5.** Statistics of the model selection used to remove variables based on AIC. The initial maximal model contained all variables while the final model did not include the effects of artificial land cover percentage, elevation, and forest canopy cover percentage.

| Model | Resid.df | Resid.Dev | Deviance | AIC |
|---|---|---|---|---|
| maximal | 26742 | 2125.5 | | 2143.487 |
| no Artificial land cover | 26742 | 2125.5 | | 2143.487 |
| no Forest canopy | 26742 | 2125.5 | | 2143.487 |
| no Elevation | 26743 | 2126.4 | 0.947 | 2142.434 |

**Table 6.** ANOVA results of the final model between deaths (dependent variable) and WT power & environmental independent variables.

| Variable | Coefficient | Deviance | Resid. Df | Resid. Dev | P-value | Signif. |
|---|---|---|---|---|---|---|
| Intercept | -111.659 | | | | | |
| Power | 1.26 | 46.928 | 26749 | 2172.1 | <0.001 | *** |
| Dist trees | -0.01 | 10.158 | 26748 | 2161.9 | 0.001 | ** |
| Dist water | 1.78 | 5.521 | 26747 | 2156.4 | 0.02 | * |
| Aspect | 0.694 | 4.153 | 26746 | 2152.3 | 0.04 | * |
| Slope | -1.163 | 5.03 | 26745 | 2147.2 | 0.025 | * |
| Agricultural cover | 0.396 | 14.759 | 26744 | 2132.5 | <0.001 | *** |
| Natural cover | 17.345 | 6.033 | 26743 | 2126.4 | 0.014 | * |

**Table 7.** Randomizations of the variables used in variance partitioning. Variance partitioning was performed between the explanatory covariates of bat deaths in the optimal model. All variables were found to be significant after 99 randomizations.

| **Variable** | **Obs** | **Z.score** | **sig95** |
|---|---|---|---|
| Power | 0.01 | 14.48 | * |
| Dist trees | 0.00 | 2.97 | * |
| Dist water | 0.00 | 9.90 | * |
| Aspect | 0.00 | 3.03 | * |
| Slope | 0.00 | 15.43 | * |
| Agricultural cover | 0.00 | 5.94 | * |
| Natural cover | 0.01 | 12.28 | * |

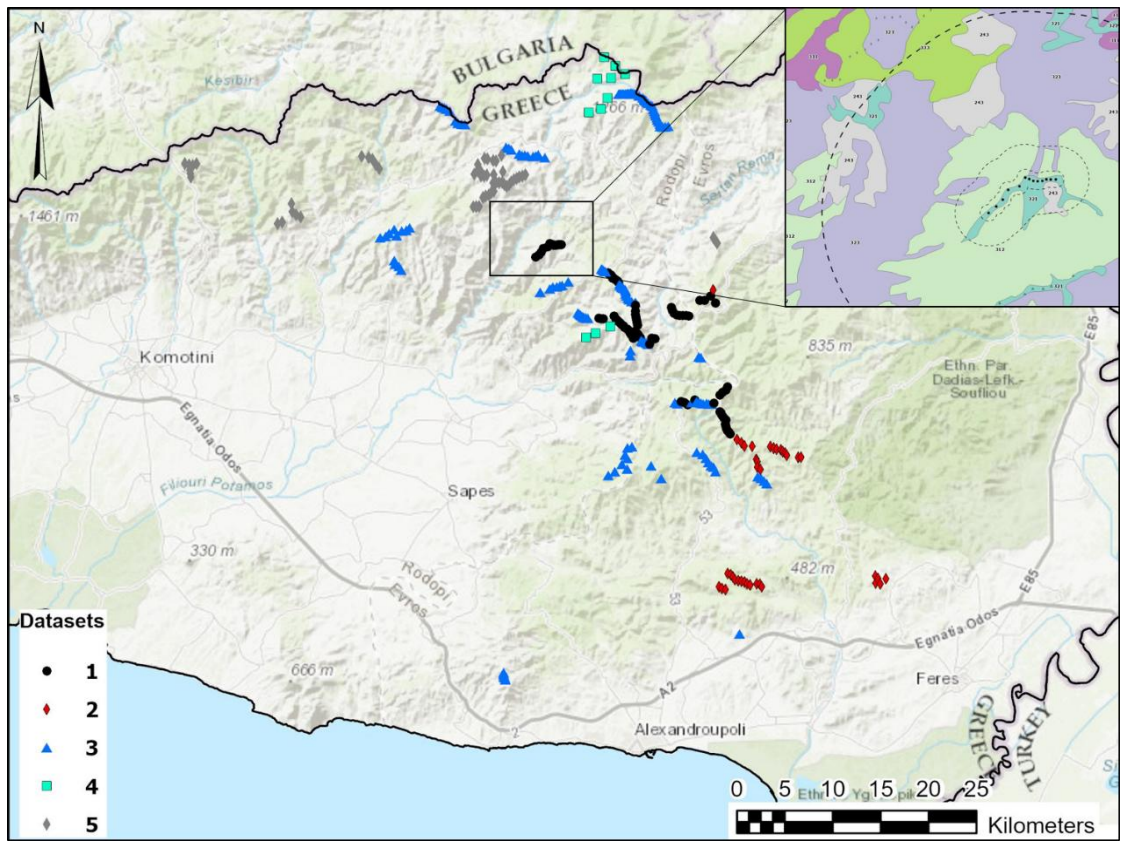

**Figure 1. a.** Overview of the geographic location of datasets 1 to 5 in Thrace region, Northern Greece. **b.** (inner figure) Detail of land cover composition at scales of 250, 1000 and 5000 m (dashed outlines) around the WTs of Monastiri area (dataset 1, black dots). Different background colours depict land cover classification types.

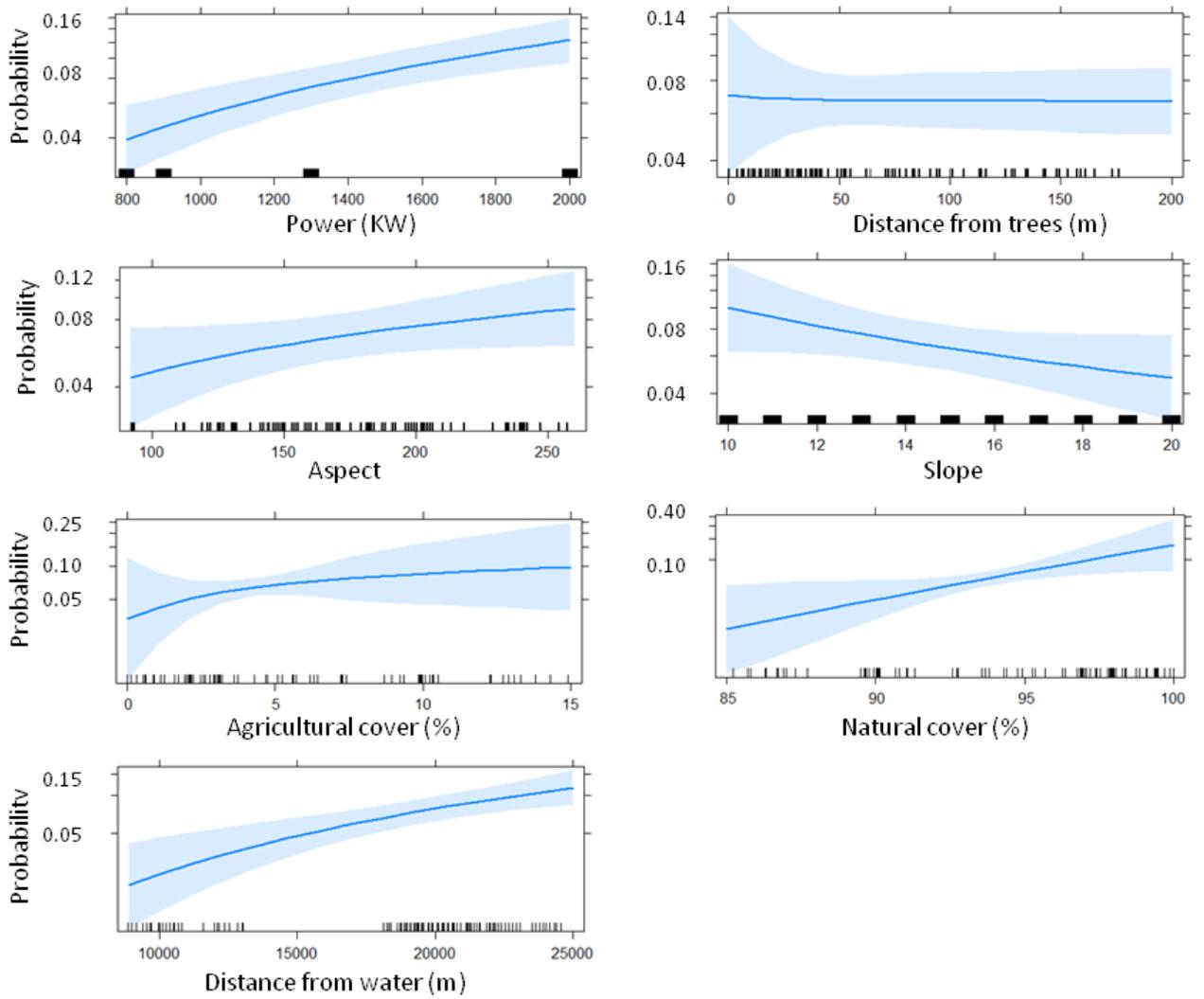

**Figure 2**. Effects of each covariate of the final model on bat deaths. Solid blue line indicates model fit, while shaded light blue envelopes indicate a 95% confidence interval. Vertical axis indicates probability of death for each variable, while horizontal axis the data range of each predictor in the final model. Thickness of black bars in the horizontal axis indicates data density.

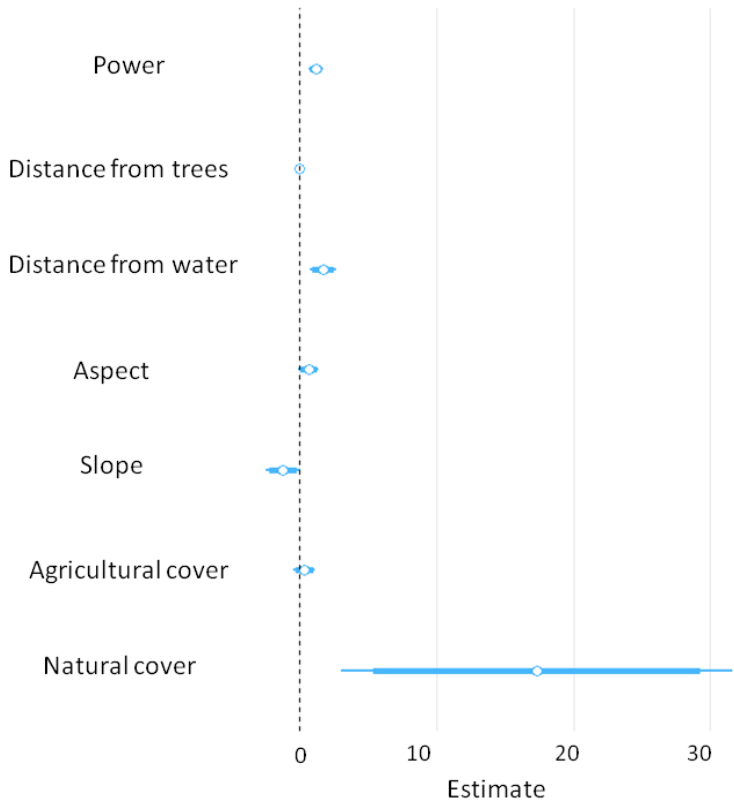

**Figure 3**. Sensitivity analysis of coefficients of the final model between bat deaths and explanatory covariates. Horizontal thin blue lines indicate a 95% confidence interval of the coefficient value of each parameter, while thick blue lines indicate a 90% confidence interval. Positive coefficient estimates indicate higher deaths while negative lower. The effects of distance from trees and agricultural land cover are crossing the dotted vertical zero line indicating that for some values their effects may not be consistent. All other effects are not crossing zero.

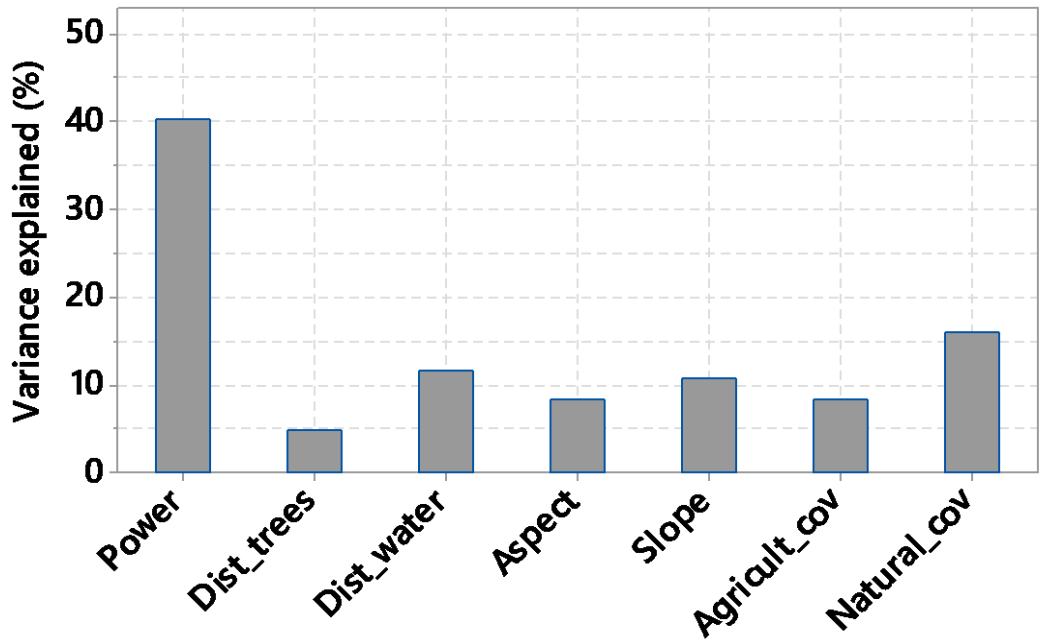

**Figure 4**. Variance partitioning of the final model covariates on bat deaths. Variance partitioning is quantifying the percentage of variance explained by each variable, and ranking their relative importance. Results add to 100%.

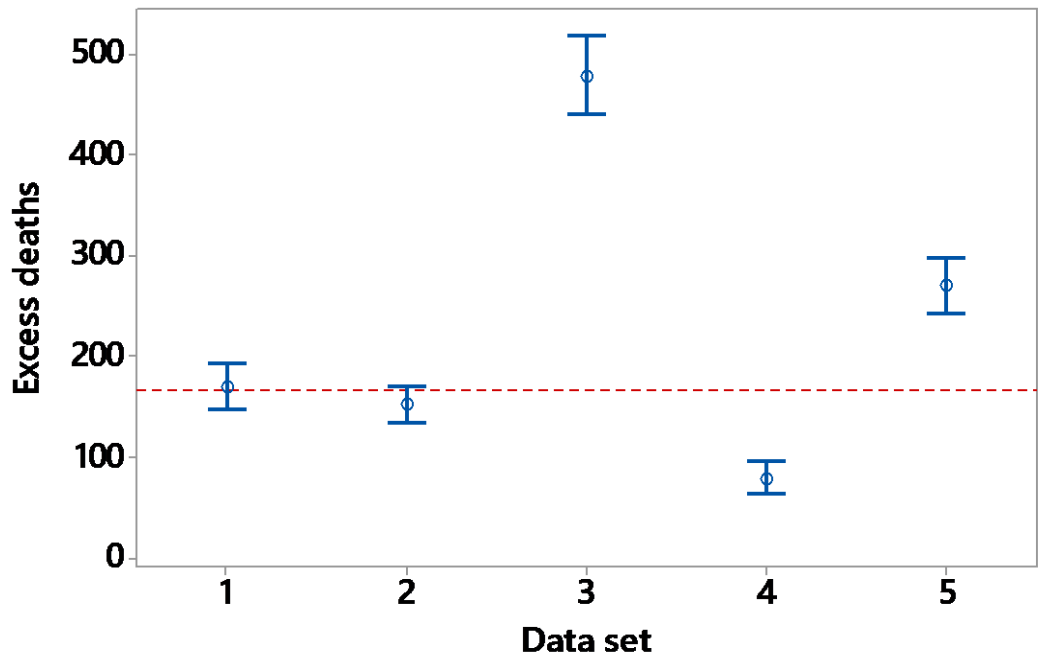

**Figure 5**. Bat deaths per year predicted in datasets 2 – 5 by the trained optimal model in dataset 1. Circles indicate the mean while whiskers indicate a 95% confidence interval of the mean. The horizontal dotted red line indicates the actual recorded number of deaths in dataset 1 in a year.